\documentclass[journal]{IEEEtran}
\usepackage{amsmath,amsfonts,amssymb}
\usepackage{algorithmic}
\usepackage{algorithm}
\usepackage{array}
\usepackage[caption=false,font=footnotesize,labelfont=sf,textfont=sf]{subfig}
\usepackage{textcomp}
\usepackage{stfloats}
\usepackage{url}
\usepackage{verbatim}
\usepackage{graphicx}
\usepackage{svg}
\usepackage{booktabs}
\usepackage{cite}
\usepackage{color}
\usepackage{multirow}

\graphicspath{{./figures/}}
\usepackage{hyperref}

\begin{document}

\title{Reinforcement Learning-Based Energy Management for Industrial Park with Heterogeneous Batteries under Demand Response}

\author{Meng Yuan,~\IEEEmembership{Member,~IEEE,} Tinghui Yan, Zhezhuang Xu,~\IEEEmembership{Member,~IEEE}
\thanks{This work was supported in part by the Marie Sk\l{}odowska-Curie Actions Postdoctoral Fellowships under the Horizon Europe programme (Grant No.~101110832) and in part by the National Natural Science Foundation of China (Grant No.~62573128). \emph{(Corresponding author: Zhezhuang Xu.)}}

 \thanks{
Meng Yuan, Tinghui Yan, and Zhezhuang Xu are with the College of Electrical Engineering and Automation, Fuzhou University, Fuzhou 350108, China. (E-mail: \tt\small zzxu@fzu.edu.cn)}} 

\markboth{ }%
{Yuan \MakeLowercase{\textit{et al.}}: RL-Based Energy Management for Industrial Park}


\maketitle

\begin{abstract}
The integration of photovoltaic (PV) systems, stationary energy storage systems (ESSs), and electric vehicles (EVs) alongside demand response (DR) programmes in industrial parks presents opportunities to reduce costs and improve renewable energy utilisation. Coordinating these resources is challenging because office and production zones have distinct operational objectives, and battery ageing costs are often ignored. This paper proposes a DR-based energy management framework that jointly optimises grid interaction costs, thermal comfort, EV departure state-of-charge requirements, carbon emissions, and battery ageing. We model heterogeneous load characteristics using a dynamic energy distribution ratio and incorporate dispatch-level ageing models for both ESS and EV batteries. The problem is formulated as a Markov decision process (MDP) and solved with a deep deterministic policy gradient (DDPG) algorithm. High-fidelity simulations using data from a practical industrial park in China show the framework maintains indoor comfort while significantly reducing total operating costs, yielding savings of 44.58\% and 40.68\% compared with a rule-based DR strategy and a conventional time-of-use arbitrage approach, respectively.
\end{abstract}


\section{Introduction}

\IEEEPARstart{T}{he} building sector, encompassing both residential and industrial buildings, is a major contributor to global energy consumption and carbon emissions. In Europe, buildings account for approximately 40\% of total regional energy consumption \cite{blomqvist2022understanding}. The overall energy efficiency of conventional buildings remains low, and achieving emission reductions in this sector is vital for meeting climate targets and ensuring energy security. To reduce reliance on fossil fuels, renewable energy sources that are increasingly abundant and cost-competitive have attracted growing attention \cite{2019Terawatt}. The installation of rooftop photovoltaic (PV) systems enables buildings to transition from passive energy consumers to active energy producers, offering a viable pathway toward nearly zero-energy buildings \cite{ascione2019building}. However, PV generation is inherently intermittent and uncertain due to diurnal, seasonal, and meteorological variations, frequently resulting in a temporal mismatch between peak PV output and peak building energy demand, thereby posing challenges to both grid stability and building energy autonomy \cite{yin2020impacts}.

To mitigate this mismatch, an increasing number of buildings have deployed battery energy storage systems (ESSs) \cite{datta2021review}. ESSs store surplus electricity during periods of excess generation and discharge it during supply shortfalls, significantly enhancing a building's solar self-consumption rate and energy resilience \cite{hannan2021battery}. In \cite{selvaraj2023smart}, an artificial intelligence-based building energy monitoring and management scheme was proposed to optimise energy consumption and promote renewable energy utilisation, however, this study did not explicitly quantify battery ageing costs and offered only limited consideration of occupant comfort. Similarly, the authors in \cite{thirugnanam2022energy} developed a reconfigurable hybrid AC/DC micro-grid architecture with an energy management strategy to minimise commercial building electricity costs and improve supply reliability, yet battery ageing costs were not incorporated.

Overall, the majority of building-side energy management system (EMS) and micro-grid studies have not systematically characterised battery ageing costs, nor have they addressed the coordinated optimisation of differentiated objectives \cite{yuan2026reinforcement}, such as office-zone comfort and production-zone carbon emissions, within industrial park buildings. Furthermore, directly employing electrochemical mechanism models involving partial differential equations and fast time-scale dynamics is impractical for dispatch-level applications, while overly simplified models fail to support accurate decision-making \cite{kuboth2019economic}. The recent widespread adoption of electric vehicles (EVs) and vehicle-to-grid (V2G) technology has broadened the scope of building-side energy storage, allowing these vehicles to function as mobile units for energy dispatch \cite{yilmaz2012review, zhang2024multi}. Given the complementary availability profiles of ESSs and EVs, coordinated dispatch strategies that integrate both resources are regarded as a promising direction for enhancing building energy system performance \cite{vanlalchhuanawmi2024energy,liu2026deep}.

Industrial parks typically encompass multiple types of loads. Conventional independent load management lacks coordination, whereas aggregated management can achieve 5\% to 6\% energy demand savings \cite{2017Conceptual, 2004Benefits}. However, integrating overall park resources among diverse loads remains a significant challenge. Office zones impose strict requirements on indoor thermal comfort, making heating, ventilation, and air conditioning (HVAC) load adjustments highly constrained \cite{tang2019model}. Conversely, production zones must maintain production efficiency while pursuing carbon reduction targets \cite{dey2023microgrid}. Balancing these divergent objectives becomes particularly complex when implementing demand response (DR) programmes, which typically guide users to adjust consumption patterns \cite{lee2013assessment}. Because directly curtailing manufacturing loads is impractical, industrial buildings can discharge battery systems to alter their net grid power exchange. This approach enhances grid stability \cite{siano2014demand}, provides economic incentives \cite{gellings2020smart}, and facilitates renewable energy integration \cite{palensky2011demand}.

Coordinating this DR strategy alongside ESSs and EVs introduces further complexity. The EMS must jointly optimise economic, comfort, low-carbon, and battery degradation objectives \cite{hossain2023review}. Current research has explored multiple technical pathways to address this. In \cite{schmitt2023regression}, hierarchical model predictive control (MPC) was combined with data-driven error compensation. Additionally, a real-coded genetic algorithm (RCGA)-based electro-thermal co-optimisation method was proposed in \cite{khan2022modeling}. However, these model-based methods face challenges in complex and uncertain scenarios because their performance degrades significantly when model mismatches or prediction errors occur. In contrast, learning-based algorithms such as reinforcement learning (RL) adaptively adjust decisions through continuous environment interaction, offering a promising alternative for real-time EMS dispatch that satisfies both carbon reduction targets and comfort boundaries \cite{2024Demand}.

Based on the current research limitations, this paper investigates the energy management dispatch optimisation problem for industrial park buildings equipped with PV systems, ESSs, EVs, and diversified loads, under a practical DR policy framework. The objective is to achieve comprehensive optimisation among system economics, low-carbon targets, and battery ageing costs, subject to DR policy requirements, comfort constraints, EV departure state-of-charge requirements, and carbon reduction targets. This presents significant challenges. First, accurate yet computationally efficient models must be constructed for the heterogeneous battery dynamics, the comfort-related loads of office zones, and the battery ageing process. Second, effective trade-offs must be achieved among demand response revenues, battery lifetime losses, comfort assurance, EV mobility needs, and carbon reduction targets. Third, energy dispatch decisions must possess multi-time-scale look-ahead optimisation capabilities.

To address these challenges, this paper proposes an energy management method based on the DDPG algorithm combined with an applicable battery ageing model, within the DR framework. The algorithm leverages real-time observations to make coordinated charge and discharge decisions for ESS and EV, incorporating comfort metrics, carbon emission intensity, and EV departure state-of-charge as key features of the state space and core weights of the reward function, thereby achieving multi-objective optimal dispatch while satisfying DR policies, load constraints, comfort requirements, EV mobility needs, and carbon reduction targets. To the best knowledge of the authors, this is the first work that jointly addresses DR participation, heterogeneous battery ageing, thermal comfort, and EV departure SoC requirements within a unified deep reinforcement learning framework for industrial park energy management. The principal contributions are as follows:

\begin{itemize}
\item A coordinated scheduling model is developed for heterogeneous loads in industrial park buildings, and a dynamic energy distribution ratio is introduced and adaptively optimised by the DDPG agent to allocate battery energy between office and production loads.
\item A heterogeneous battery model is established for stationary ESS and EV systems, and a dispatch-level battery ageing cost formulation is incorporated into the energy management problem.
\item A practical demand response based energy management problem is formulated to jointly consider grid interaction cost, thermal comfort, EV departure SoC requirement, carbon emissions, and battery ageing, and it is solved using a deep reinforcement learning method.
\end{itemize}

The remainder of this work is organised as follows. Section~\ref{sec:Problem_formulation} presents the system model and problem formulation. Section~\ref{sec:Proposed method} details the methodologies and the DDPG-based energy management algorithm. Section~\ref{sec:Results} verifies the effectiveness of the proposed approach through simulations based on practical data. Finally, Section~\ref{sec:Conclusion} summarises the findings.

\section{System Models and Problem Formulation}\label{sec:Problem_formulation}

\begin{figure}[tb]
    \centering
    \includegraphics[width=0.85\linewidth]{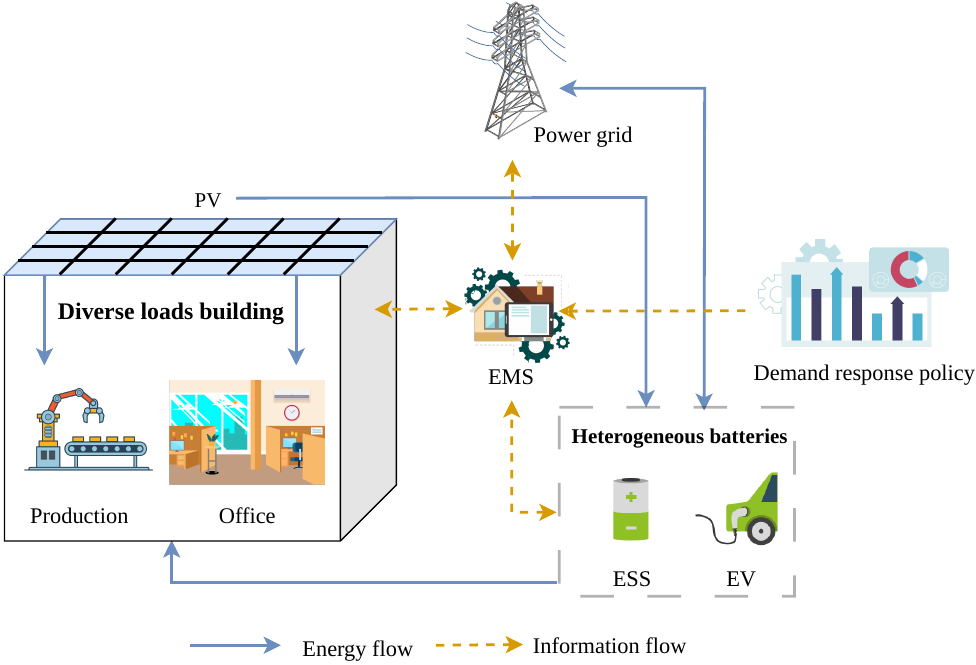}
    \caption{Schematic diagram of the industrial park energy management system.}\label{fig:schematic}
\end{figure}

This study considers an energy optimisation framework for an industrial park building, as illustrated in Fig.~\ref{fig:schematic}. The system integrates diverse loads, such as production and office loads, coupled with solar power generation and heterogeneous batteries, specifically ESSs and EVs. Notably, these heterogeneous batteries are modelled with distinct characteristics. The system operates in discrete time slots indexed by $t \in \{1, 2, \ldots, T_{\mathrm{total}}\}$, with a sampling interval of $\Delta t = 1$~hour.

In this section, the models of different system components are described first, followed by the problem formulation considering the demand response policy.

\subsection{System Modelling}

\subsubsection{PV Model}

The power generated by the PV system at time slot $t$, denoted as $P^{\text{PV}}(t)$, is constrained by the available solar irradiance and the capacity of the installed panels. In our energy dispatch model, we assume the photovoltaic output is deterministic, given the maturity of PV forecasting. The operational constraint for the PV system is given by
\begin{equation}
  0 \le P^{\text{PV}}(t) \le P^{\text{PV}}_{\max},
  \label{eq:pv_model}
\end{equation}
where $P_{\max}^{\text{PV}}$ represents the installed capacity of the PV panels. The generated power $P^{\text{PV}}(t)$ contributes to the building's energy supply and is managed by the EMS to either serve the building loads or be stored in the ESS.

\subsubsection{HVAC Model}

As the largest contributor to energy consumption in office zones, HVAC systems can be dynamically adjusted to maintain thermal comfort. While true thermal comfort depends on numerous factors such as average radiation temperature, relative humidity, air velocity, clothing insulation, and metabolic rate, incorporating all these variables creates a highly complex system representation. To maintain a tractable energy dispatch model, we use indoor air temperature as a practical proxy for thermal comfort. The building thermal dynamics are therefore described by the following simplified first-order model \cite{lu2020optimal}:
\begin{equation}
\begin{split}
T_{\mathrm{in}}(t+1) =& \varepsilon T_{\mathrm{in}}(t) + (1-\varepsilon) \\
&\times \left(T_{\mathrm{out}}(t) + \delta(t)\frac{\eta^{\mathrm{HVAC}} P^{\mathrm{HVAC}}(t)}{G_{\text{b}}}\right),
\end{split}
\label{eq:temperature}
\end{equation}
where $T_{\mathrm{out}}(t)$ is the outdoor temperature, $G_{\text{b}}$ is the building's thermal conductance, $\eta^{\mathrm{HVAC}}$ is the efficiency ratio, and $\varepsilon$ is the thermal inertia factor. The operating mode of the system is determined by \( \delta(t) \), defined as:
\begin{equation}
\delta(t) =
\begin{cases}
+1 & \text{heating mode}, \\
-1 & \text{cooling mode}, \\
0 & \text{HVAC off}.
\end{cases}
\end{equation}
Let \( P^{\text{HVAC}}(t) \geq 0 \) denote the magnitude of the power input. The effective electrical power is given by \( \delta(t) \cdot P^{\text{HVAC}}(t) \), and the power constraint is
\begin{equation}
0 \leq P^{\text{HVAC}}(t) \leq P^{\text{HVAC}}_{\max},
\end{equation}
where \( P^{\text{HVAC}}_{\max} \) is the rated power of the HVAC system.

To measure thermal comfort, a comfortable temperature range is used as a representation of thermal comfort, given by
\begin{equation}
{T}_{\min} \leq {T}_{\text{in}}(t) \leq {T}_{\max},
\end{equation}
where ${T}_{\text{in}}(t)$ is the indoor temperature, and $T_{\min}$ and $T_{\max}$ are the minimum and maximum comfort levels, respectively.



\subsubsection{Battery Model}

Let ${\text{SoC}}^{j}(t)$ be the stored energy of the battery at time slot $t$, where $j \in\{ \text{ESS}, \text{EV}\}$ represents ESS or EV. The energy storage dynamic model of the battery is established as \cite{cao2020deep}

\begin{equation}
\text{SoC}^{j}(t+1) =
\begin{cases}
\text{SoC}^{j}(t) + \dfrac{P^{j}(t)}{E^{j}} \left( \dfrac{s^j_{\text{d}}}{\eta^j_{\text{d}}} + s^j_{\text{c}} \eta^j_{\text{c}} \right),&  P^{j}(t) \neq 0 \\
\text{SoC}^{j}(t)-\dfrac{P_{\text{standby}}^{j}(t)}{E^{j}},&  P^{j}(t) = 0
\end{cases}
\end{equation}
where $\text{SoC}^j(t)$ is the state of charge at time slot $t$, $P^{j}(t)$ is the output power (with $P^j(t) > 0$ for charging and $P^{j}(t) < 0$ for discharging), $P^{j}_{\text{standby}}(t)$ is the standby loss of the battery when it is idle at time slot $t$, $E^{j}$ denotes the battery's total energy capacity, $s^j_{\text{d}}$ and $s^j_{\text{c}}$ are binary variables that control the charging and discharging of the battery, and $\eta^j_{\text{c}}$ and $\eta^j_{\text{d}}$ are the charging and discharging efficiencies, respectively.

Since the battery cannot charge beyond its upper limit $\text{SoC}^{j}_{\text{max}}$ or discharge below the minimum energy level $\text{SoC}^{j}_{\text{min}}$, the following constraint applies:
\begin{equation}
\text{SoC}^j_{\min} \leq \text{SoC}^{j}(t) \leq \text{SoC}^j_{\max}.
\end{equation}

Because the charging and discharging power of the battery is limited, the power constraint is given by
\begin{equation}
-P^{j}_{\text{rated}} \leq {P}^{j}(t) \leq P^{j}_{\text{rated}},
\end{equation}
where $P^{j}_{\text{rated}}$ is the rated charging and discharging power. To avoid simultaneous charging and discharging of the battery, the following condition must hold:
\begin{equation}
s^j_{\text{d}} s^j_{\text{c}} = 0.
\end{equation}

\subsubsection{Power Balancing}
To maintain the power balance of the park building, the total power supply must equal the total power demand of the production load, office load, and EV charging. Therefore, the power balance equation is given by
\begin{equation}
P^{\text{grid}}(t) + P^{\text{PV}}(t)= P^{\text{load}}(t)+ P^{\text{HVAC}}(t) + P^{\text{ESS}}(t) + P^{\text{EV}}(t),
\label{powerblance}
\end{equation}
where $P^{\text{load}}(t) = L^{\text{pro}}(t) + L^{\text{off}}(t)$ represents the combined production load $L^{\text{pro}}(t)$ and office load $L^{\text{off}}(t)$, which is separate from the HVAC power $P^{\text{HVAC}}(t)$. The term $P^{\text{grid}}(t)$ is the power exchanged with the utility grid, where $P^{\text{grid}}(t) > 0$ indicates energy purchased from the grid and $P^{\text{grid}}(t) < 0$ represents surplus energy sold back.

\subsection{Multi-Objective Optimisation Problem Formulation}

The EMS aims to jointly maximise DR revenue while minimising grid interaction costs, carbon emissions, and battery ageing over the optimisation horizon. This introduces several competing trade-offs. Although discharging the battery generates DR revenue and reduces grid electricity costs, each cycle accelerates degradation and incurs ageing penalties. Time-shifted arbitrage charging also increases these cycling losses. Furthermore, demand-side adjustments face operational boundaries: modulating HVAC power is constrained by predefined thermal comfort limits, EV batteries must retain a sufficient state of charge by the departure time to satisfy mobility needs, and battery dispatching to meet DR targets must be balanced against the corresponding degradation costs. Finally, key system parameters such as PV generation, building load demand, outdoor temperature, and EV availability are inherently stochastic. This uncertainty must be accounted for in the optimisation objective.

To address these challenges, we reformulate the building energy management task as a Markov decision process (MDP) and solve it using the DDPG algorithm described in Section~\ref{sec:Proposed method}. Based on the above system models, the multi-objective optimisation problem investigated in this work is formally stated as
\begin{equation}
\begin{aligned}
\max \sum_{t=1}^{T_{\text{total}}} \mathbb{E} \bigg\{
& R_{\text{dr}}(t) - C_{\text{grid}}(t) - C_{\text{COx}}(t) \\
&- C_{\text{deg}}(t) - C_{\text{T}}(t) - C_{\text{SoC}}(t) \bigg\}
\end{aligned}
\end{equation}

\begin{equation}
\text{s.t.} \quad (\ref{eq:pv_model})-(\ref{soc_penalty}),
\end{equation}
where $R_{\text{dr}}(t)$ is the demand response revenue, $C_{\text{grid}}(t)$ is the grid interaction cost, $C_{\text{COx}}(t)$ is the carbon emission cost, $C_{\text{deg}}(t)$ is the battery ageing cost, $C_{\text{T}}(t)$ is the temperature deviation penalty, and $C_{\text{SoC}}(t)$ is the SoC departure penalty. Each cost term is detailed in the following.

\subsubsection{Demand Response Revenue}

The demand response revenue $R_{\text{dr}}(t)$ is modelled according to the Fujian Electric Power Demand Response Implementation Plan issued in July 2024\cite{fujian2024demandresponse}. Under this policy, the grid operator sends day-ahead dispatching signals requesting participating enterprises to reduce their apparent grid consumption during peak periods. Enterprises that successfully lower their demand are compensated at a pre-agreed declared unit price $p_{\text{dr}}$, with the compensation amount determined by the magnitude and ratio of the achieved load reduction. Because directly curtailing production processes is often impractical, the industrial park instead discharges its batteries to offset grid consumption, thereby fulfilling the DR obligation without disrupting operations.

A critical element of the policy is the \emph{baseline load}, which serves as a reference to quantify the actual load reduction achieved by each participant and to prevent unintended subsidies from natural load declines. The baseline load is computed as the average production load over the preceding $N$ similar days. Specifically, $N=5$ for working days and $N=3$ for non-working days, reflecting the distinct consumption patterns across day types. Let $h \in \{1, \ldots, H\}$ denote the intra-day hour index, where $H = 24$ for hourly sampling. The baseline load at hour $h$ is then given by
\begin{equation}
L^{\text{base}}(h) = \frac{1}{N} \sum_{i=1}^{N} {L}_{i}^{\text{pro}}(h),
\end{equation}
where ${L}_{i}^{\text{pro}}(h)$ denotes the production load at hour $h$ on the $i$-th preceding similar day in kWh. This slot-by-slot averaging ensures that the baseline at each hour of the DR day represents the typical consumption at that specific hour, rather than a single daily total. For notational convenience, we write $L^{\text{base}}(t) \triangleq L^{\text{base}}(\varphi(t))$ in all subsequent equations, where $\varphi(t) = \bigl((t-1) \bmod H\bigr) + 1$ maps the global time index $t$ to the corresponding intra-day hour.

The total adjusted load $\Delta L(t)$, which quantifies the effective contribution to the DR event, is constructed from four components: the deviation between the baseline and the actual production load, the PV energy generated, and the battery energy allocated to the production zone. It is defined as
\begin{equation}
\Delta {L}(t) = {L}^{\text{base}}(t) - {L}^{\text{pro}}(t) + {Q}^{\text{PV}}(t) + {Q}^{\text{pro}}(t),
\end{equation}
where the energy quantities involved are given by
\begin{equation}
{Q}^{\text{PV}}(t) = {P}^{\text{PV}}(t) \, \Delta t,
\end{equation}
\begin{equation}
{Q}^{\text{bat}}(t) = \bigl({P}^{\text{ESS}}(t) + {P}^{\text{EV}}(t)\bigr) \, \Delta t,
\end{equation}
\begin{equation}
{Q}^{\text{pro}}(t) = r_{\text{dis}}(t) \, {Q}^{\text{bat}}(t).
\end{equation}
Here, ${Q}^{\text{PV}}(t)$ is the PV energy output, ${Q}^{\text{bat}}(t)$ is the total battery energy dispatched by both the ESS and EV, $r_{\text{dis}}(t)$ is the dynamic energy distribution ratio that allocates battery energy between production and office loads, and ${Q}^{\text{pro}}(t)$ is the portion of battery energy allocated to the production load.

The grid operator evaluates each enterprise's contribution through the response ratio $r_{\text{dr}}(t)$, defined as the proportion of the actual load adjustment relative to the invited response target $L_{\text{dr}}$:
\begin{equation}
r_{\text{dr}}(t) = \frac{\Delta {L}(t)}{L_{\text{dr}}},
\end{equation}
where $L_{\text{dr}}$ is the invited response load in kWh. The DR revenue is then determined by a tiered compensation structure that rewards higher response ratios:
\begin{equation}
R_{\text{dr}}(t) =
\begin{cases}
0 & \text{if } r_{\text{dr}}(t) < 0.5, \\
0.6 \, p_{\text{dr}} \, \Delta {L}(t) & \text{if } 0.5 \leq r_{\text{dr}}(t) \leq 0.8, \\
p_{\text{dr}} \, \Delta {L}(t) & \text{if } 0.8 < r_{\text{dr}}(t) \leq 2.0, \\
2 \, p_{\text{dr}} \, \Delta {L}(t) & \text{otherwise},
\end{cases}
\end{equation}
where $p_{\text{dr}}$ is the declared unit compensation price for DR participation in RMB/kWh. When the response ratio falls below 0.5, the enterprise receives no compensation. Partial compensation at 60\% of the unit price applies when the ratio lies between 0.5 and 0.8. Full compensation is granted for ratios between 0.8 and 2.0. It is worth noting that the above DR revenue model is entirely based on the official Fujian provincial DR implementation plan, thereby ensuring that the proposed framework can be directly applied to real-world energy management scenarios.

\subsubsection{Grid Interaction Cost}
The grid interaction cost $C_{\text{grid}}(t)$ represents the net cost of energy exchanged between the building and the utility grid. Specifically,
\begin{align}
C_{\text{grid}}(t) &= Q^{\text{grid}}(t) \, p(t) \notag \\
&= P^{\text{grid}}(t) \, \Delta t \, p(t),
\end{align}
where ${Q}^{\text{grid}}(t)$ is the amount of electricity exchanged between the building and the utility grid at time slot $t$, and $p(t)$ is the electricity price.

\subsubsection{Carbon Emission Cost}

Beyond direct electricity expenses, carbon emissions from building loads increasingly constitute a direct financial liability. With the widespread implementation of carbon pricing mechanisms such as carbon taxes and emission trading schemes, carbon output is now directly monetised. In this work, the carbon emission cost of the building $C_{\text{COx}}(t)$ is defined as
\begin{equation}
C_{\text{COx}}(t)=\rho_{\text{c}} \, \omega \, [Q^{\text{grid}}(t)]^{+},
\end{equation}
where $\rho_{\text{c}}$ is the carbon tax rate, $\omega$ is the carbon emission intensity, and $[\cdot]^{+} = \max(0,\,\cdot)$ denotes the positive part operator. The term $[Q^{\text{grid}}(t)]^{+}$ ensures that only electricity purchased from the grid incurs a carbon cost, since PV generation is carbon-free.

\subsubsection{Battery Ageing Cost}
When formulating energy management strategies, battery ageing is often overlooked, leading to inaccurate cost estimates. To address this, we incorporate dispatch-level ageing models for both battery types in this work. As introduced in Section~\ref{sec:Problem_formulation}, the ESS employs $\mathrm{LiFePO_4}$ (LFP) cells, which offer higher safety and longer cycle life, while the EV battery uses $\mathrm{Li(NiMnCo)O_{2}}$ (NMC) cells, which provide higher energy density but exhibit shorter lifespan. Since the two chemistries degrade through different mechanisms, separate empirical ageing models are adopted.

The cycling capacity loss of the LFP battery is modelled as \cite{naumann2020analysis}
\begin{equation}
Q_{\text{loss}}^{\text{LFP}} = k_{\text{C}}(C_{\text{rate}}) \, k_{\text{D}}(\text{DOD}) \, {N_\text{EFC}}^{Z_{\text{cyc}}},
\end{equation}
\begin{equation}
k_{\text{C}}(C_{\text{rate}}) = a_{1} \, C_{\text{rate}} + a_{2},
\end{equation}
\begin{equation}
k_{\text{D}}(\text{DOD}) = a_{3} \, (\text{DOD} - 0.6)^3 + a_{4},
\end{equation}
where $Q_{\text{loss}}^{\text{LFP}}$ is the fractional capacity loss, $N_{\text{EFC}}$ is the equivalent full cycle count, $\text{DOD}$ is the depth of discharge, $C_{\text{rate}}$ is the charging or discharging C-rate, $k_{\text{C}}$ and $k_{\text{D}}$ are the C-rate and DOD stress factors, respectively, $Z_{\text{cyc}}$ is the cycling ageing exponent, and $a_{1}$ to $a_{4}$ are empirical fitting parameters listed in Table~\ref{tab:cycle_aging}.

\begin{figure}[tb]
    \centering
    \includegraphics[width=0.8\linewidth]{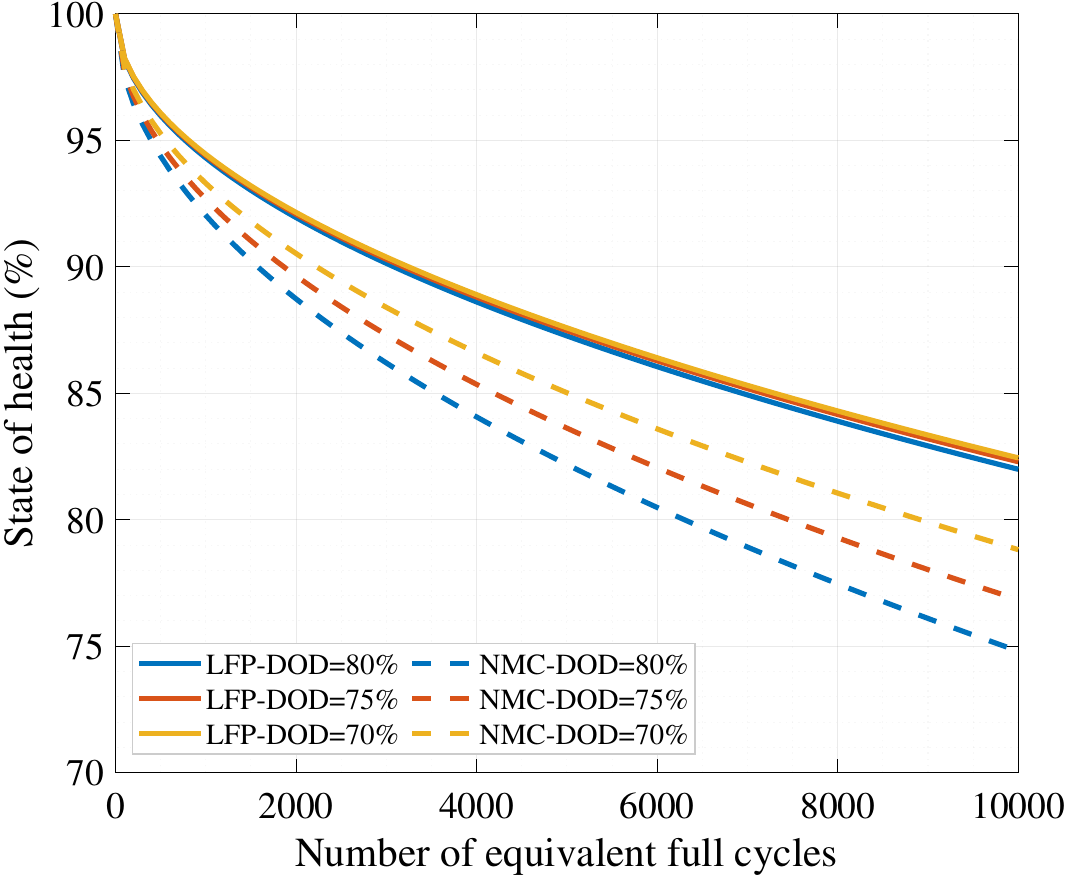}
    \caption{Cycling ageing results of LFP/NMC batteries evaluated at 40$^{\circ}\mathrm{C}$ and a C-rate of 1.0.}\label{fig:battery_aging}
\end{figure}

The cycling capacity loss of the NMC battery is modelled as \cite{schmalstieg2014holistic}
\begin{equation}
Q_{\text{loss}}^{\text{NMC}} = b_{\text{cap}}(\text{DOD}) \sqrt{Q(\text{DOD})},
\end{equation}
\begin{equation}
b_{\text{cap}}(\text{DOD}) = b_{1} (V_{\text{avg}} - V_{0})^{2} + b_{2} + b_{3} \, \text{DOD},
\label{eq:empirical_formula}
\end{equation}
\begin{equation}
Q(\text{DOD}) = N_{\text{EFC}} \, C_{0} \, \text{DOD},
\label{eq:Q_formula}
\end{equation}
where $Q_{\text{loss}}^{\text{NMC}}$ is the fractional capacity loss, $b_{\text{cap}}$ is the capacity degradation coefficient, $V_{\text{avg}}$ is the average operating voltage, $V_{0}$ is the reference voltage, $C_0$ is the initial cell capacity, and $b_{1}$ to $b_{3}$ are empirical fitting parameters listed in Table~\ref{tab:cycle_aging}. Fig.~\ref{fig:battery_aging} illustrates the ageing behaviour of both battery types under different operating conditions.

\begin{table}[tb] 
\centering
\caption{Cycle ageing model parameters of batteries}
\label{tab:cycle_aging}
\begin{tabular}{lclc}
\toprule
\multicolumn{2}{c}{LFP} & \multicolumn{2}{c}{NMC} \\
\cmidrule(r){1-2} \cmidrule(l){3-4}
Parameter & Value & Parameter & Value \\
\midrule
$a_1$ & 0.0630 & $b_1$ & $7.348\times10^{-3}$ \\
$a_2$ & 0.0971 & $b_2$ & $7.60\times10^{-4}$ \\
$a_3$ & 4.0253 & $b_3$ & $4.081\times10^{-3}$ \\
$a_4$ & 1.0923 & $V_0$ & 3.667 \\
$Z_{\text{cyc}}$ & 0.5 & $V_{\text{avg}}$ & 3.7 \\
& & $C_0$ & 2.05 \\
\bottomrule
\end{tabular}
\end{table}

To translate the above capacity loss into a monetary cost at dispatch level, inspired by \cite{cao2020deep}, we define the ageing cost coefficient $\alpha^j_d$ as the cost incurred per kWh of energy throughput:
\begin{equation}
	\alpha^j_{d} = \frac{\Delta Q^j_{\text{loss}} \, E^{j}}{\sum_{t=1}^{T_{\mathrm{cyc}}} |P^j(t)|} \, C^j_{\text{kWh}},
\end{equation}
where $\Delta Q^j_{\text{loss}}$ is the fractional capacity loss of battery $j$ over one representative cycle of duration $T_{\mathrm{cyc}}$, $E^{j}$ is the rated energy capacity, and $C^j_{\text{kWh}}$ is the procurement cost of the battery per kWh in RMB/kWh. The battery ageing cost at each time step is then given by
\begin{equation}
C^j_{\text{deg}}(t) = \alpha^j_d \, |P^j(t)| \, \Delta t, \quad j \in \{\text{ESS}, \text{EV} \}.
\label{aging_cost}
\end{equation}

\subsubsection{Temperature Deviation Penalty}
In order to ensure that the indoor temperature can meet the requirements of building comfort, a penalty term $C_{\text{T}}(t)$ is introduced, which is defined as
\begin{equation}
C_{\text{T}}(t) = [T_{\text{in}}(t)-T_{\max}]^{+} +  [T_{\min}-T_{\text{in}}(t)]^{+},
\label{temperture_penalty}
\end{equation}
where $[\cdot]^{+} = \max(0, \cdot)$ denotes the positive-part operator. During training, a tighter temperature range than the nominal comfort bounds is imposed to reduce temperature fluctuations near the boundaries.

\subsubsection{SoC Departure Constraint Penalty}

To ensure that the EV SoC can meet the driving demand of electric vehicles when they leave the site after participating in energy management, a penalty $C_{\text{SoC}}(t)$ is introduced, which is defined as
\begin{equation}
C_{\text{SoC}}(t) = [\text{SoC}_{\text{lim}}-\text{SoC}^{\text{EV}}(t)]^{+},
\label{soc_penalty}
\end{equation}
where $\text{SoC}_{\text{lim}}$ is the minimum acceptable SoC at the EV departure time.

\section{Proposed Method}\label{sec:Proposed method}

In this work, the energy management problem is formulated as an MDP and solved using a model-free deep reinforcement learning algorithm. Because the SoC of each battery at the next time slot depends only on its current SoC and the action taken, the energy management problem naturally satisfies the Markov property and can therefore be cast as an MDP.

\begin{figure}[tb]
\centering
\includegraphics[width=0.85\linewidth]{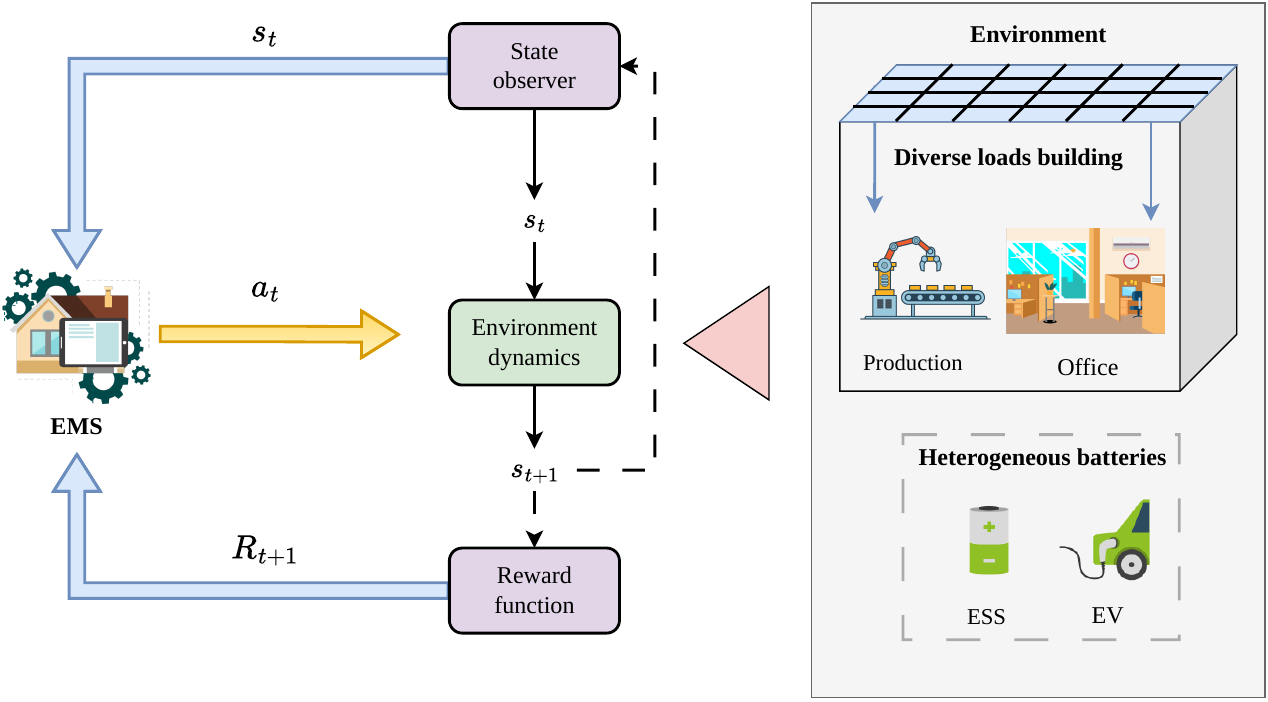}
\caption{Agent-environment interaction in the MDP.}
\label{fig:MDP}
\end{figure}

A discounted MDP is formally defined as a five-tuple $M = (\mathcal{S}, \mathcal{A}, \mathcal{P}, \mathcal{R}, \gamma)$, comprising the set of environment states $\mathcal{S}$, the set of actions $\mathcal{A}$, the transition probability function $\mathcal{P}: \mathcal{S} \times \mathcal{A} \times \mathcal{S} \rightarrow [0, 1]$, the reward function $\mathcal{R}$, and a discount factor $\gamma \in [0, 1]$. In this framework, the agent represents the learner and decision-maker, i.e., the EMS agent, which interacts with the environment comprising external elements such as photovoltaic power generation, building loads, outdoor temperature, ESS, EV, and the utility grid. This interaction is illustrated in Fig.~\ref{fig:MDP}. At each time slot, the EMS agent observes the environment state $s(t) \in \mathcal{S}$, selects an action $a(t) \in \mathcal{A}$, after which the environment transitions to a new state $s({t+1})$ and returns a corresponding reward $R({t+1})$.

In the following subsections, we elaborate on the MDP components, namely the state space, action space, and reward function, and then present the DDPG-based solution algorithm.

\subsection{MDP Formulation}

\subsubsection{Environment States}
The environment state at time slot $t$ consists of nine components: production load $L^{\text{pro}}(t)$, office load $L^{\text{off}}(t)$, demand response baseline load $L^{\text{base}}(t)$, outdoor temperature $T_{\mathrm{out}}(t)$, indoor temperature $T_{\mathrm{in}}(t)$, PV energy output $Q^{\text{PV}}(t)$, electricity price $p(t)$, ESS state of charge $\text{SoC}^{\text{ESS}}(t)$, and EV state of charge $\text{SoC}^{\text{EV}}(t)$. The state vector is denoted as
\begin{equation}
\begin{split}
s(t) = \bigl\{ & L^{\text{pro}}(t),\, L^{\text{off}}(t),\, L^{\text{base}}(t),\, T_{\mathrm{out}}(t),\, T_{\mathrm{in}}(t), \\
               & Q^{\text{PV}}(t),\, p(t),\, \text{SoC}^{\text{ESS}}(t),\, \text{SoC}^{\text{EV}}(t) \bigr\}.
\end{split}
\end{equation}

\subsubsection{Action}
The EMS agent determines four continuous control variables at each time slot: the charging/discharging power of the ESS and EV, the energy distribution ratio between production and office loads, and the HVAC operating power. Once the battery and HVAC powers are determined, the grid exchange power $P^{\text{grid}}(t)$ follows directly from the power balance equation~(\ref{powerblance}). Each control variable is represented by a continuous signal in $[-1,1]$, which the EMS maps to the corresponding physical power range. This unified representation naturally encompasses charging, discharging, heating, cooling, and standby modes without requiring separate binary variables. The action vector is
\begin{equation}
a(t) = \{P^{\text{ESS}}(t),\, P^{\text{EV}}(t),\, r_{\text{dis}}(t),\, P^{\text{HVAC}}(t)\}.
\end{equation}

In addition to the rated power constraint defined in Section~\ref{sec:Problem_formulation}, the actual charging and discharging power must respect the SoC limits. Recall that $E^j$, $\eta^j_{\text{c}}$, $\eta^j_{\text{d}}$, $\text{SoC}^j_{\min}$, and $\text{SoC}^j_{\max}$ are defined in the battery model of Section~\ref{sec:Problem_formulation}. From the SoC dynamics, the charging power is bounded by the remaining capacity up to $\text{SoC}^j_{\max}$, and the discharging power is bounded by the available energy above $\text{SoC}^j_{\min}$:
\begin{equation}
0 \leq P^j(t) \leq \frac{\bigl(\text{SoC}^j_{\max}-\text{SoC}^j(t)\bigr) \, E^{j}}{\eta^j_{\text{c}}},
\label{eq:charge_bound}
\end{equation}
\begin{equation}
\bigl(\text{SoC}^j_{\min}-\text{SoC}^j(t)\bigr) \, E^{j} \, \eta^j_{\text{d}} \leq P^j(t) \leq 0.
\label{eq:discharge_bound}
\end{equation}

\subsubsection{Reward Function}

At each time slot, the agent executes action $a(t)$, which causes the environment to transition from state $s(t)$ to $s(t+1)$ and yields a scalar reward $R(t)$. The reward is designed so that maximising the cumulative reward aligns with the optimisation objective defined in Section~\ref{sec:Problem_formulation}. It comprises six components: demand response revenue, grid interaction cost penalty, battery ageing penalty, carbon emission penalty, temperature deviation penalty, and EV SoC departure penalty. The composite reward function is defined as
\begin{equation}
\begin{split}
R(t) &= \lambda_1 R_{\text{dr}}(t) - \lambda_2 C_{\text{grid}}(t) - \lambda_3 C_{\text{T}}(t) \\
&\quad - \lambda_4 C_{\text{SoC}}(t) - \lambda_5 C_{\text{deg}}(t) - \lambda_6 C_{\text{COx}}(t),
\end{split}
\end{equation}
where $C_{\text{deg}}(t) = \sum_{j \in \{\text{ESS},\text{EV}\}} C^j_{\text{deg}}(t)$ is the total battery ageing cost, and $\lambda_1$--$\lambda_6$ are positive weighting coefficients that balance the relative importance of each objective. Among these, $R_{\text{dr}}(t)$, $C_{\text{grid}}(t)$, $C_{\text{deg}}(t)$, and $C_{\text{COx}}(t)$ correspond directly to the terms in the optimisation objective, whilst $C_{\text{T}}(t)$ and $C_{\text{SoC}}(t)$ are penalty terms that enforce thermal comfort and EV departure constraints, respectively.

\subsubsection{Action-Value Function}
The goal of the EMS agent is to maximise its expected cumulative discounted reward, defined as $G_t = \sum_{i=0}^{\infty} \gamma^{i} R(t+1+i)$. Let $Q_{\pi}(s, a)$ denote the action-value function under policy $\pi$, which maps each state--action pair to the expected return when action $a$ is taken in state $s$ and policy $\pi$ is followed thereafter. The optimal action-value function $Q^{*}(s, a) = \max_\pi Q_{\pi}(s, a)$ satisfies the Bellman optimality equation:
\begin{equation}
\begin{aligned}
Q^{*}(s, a) & = \mathbb{E}\left[R({t+1}) + \gamma \max_{a^{\prime}} Q^{*}\left(s({t+1}), a^{\prime}\right) \mid s, a\right] \\
& = \sum_{s^{\prime}, r} P\left(s^{\prime}, r \mid s, a\right)\left[r + \gamma \max_{a^{\prime}} Q^{*}\left(s^{\prime}, a^{\prime}\right)\right],
\end{aligned}
\end{equation}
where $s^{\prime} \in \mathcal{S}$, $r \in \mathcal{R}$, $a^{\prime} \in \mathcal{A}$, and $P \in \mathcal{P}$.

Solving the Bellman equation requires knowing the state transition probability $P(s^{\prime}, r \mid s, a)$, which is difficult to model accurately in building energy systems due to the stochastic nature of PV generation, occupant behaviour, and outdoor temperature. Model-free reinforcement learning avoids this difficulty by estimating $Q^*$ directly from sampled transitions. For discrete action spaces, the deep $Q$-network (DQN) algorithm uses a neural network to approximate $Q^*$. However, the action space in this problem is continuous, which precludes the $\max_{a'}$ operation in DQN. To address this, we adopt the DDPG algorithm, which maintains a separate actor network to approximate the optimal policy in continuous domains.

\subsection{DDPG-based Energy Management Algorithm}

DDPG is an off-policy actor-critic algorithm for continuous control. The actor network $\mu(s|\theta^\mu)$ parameterises a deterministic policy that maps states to actions, while the critic network $Q(s,a|\theta^Q)$ estimates the corresponding action-value function. To improve learning stability, DDPG uses experience replay to decorrelate training samples and target networks to stabilise the update process.

The actor network is updated using the deterministic policy gradient theorem. The gradient of the expected return with respect to the actor parameters is approximated as
\begin{equation}
\begin{aligned}
\nabla_{\theta^\mu} J
&\approx \mathbb{E}_{s_t \sim \rho^\beta}
\Bigl[
\nabla_a Q(s_t, a|\theta^Q)\big|_{a=\mu(s_t)} \\
&\qquad\qquad \nabla_{\theta^\mu} \mu(s_t|\theta^\mu)
\Bigr],
\end{aligned}
\label{eq:ddpg_actor_grad}
\end{equation}
where $\rho^\beta$ is the state visitation distribution under a behaviour policy $\beta$, which is typically the current policy with added exploration noise. In implementation, this expectation is estimated from minibatches sampled from the replay memory $\mathcal{D}$. The critic network $Q(s,a|\theta^Q)$ is updated by minimising the temporal-difference loss
\begin{equation}
\begin{aligned}
L(\theta^Q)
&= \mathbb{E}_{(s_t, a_t, r_t, s_{t+1}, d_t) \sim \mathcal{D}}
\Bigl[
\bigl(y_t - Q(s_t, a_t|\theta^Q)\bigr)^2
\Bigr],
\end{aligned}
\label{eq:ddpg_critic_loss}
\end{equation}
where the target value $y_t$ is given by
\begin{equation}
\begin{aligned}
y_t = r_t + \gamma (1-d_t) \, Q'\Bigl(
s_{t+1}, \, \mu'(s_{t+1}|\theta^{\mu'}) \,\big|\, \theta^{Q'}
\Bigr),
\end{aligned}
\label{eq:ddpg_target}
\end{equation}
where $Q'$ and $\mu'$ are the target critic and target actor networks, respectively, with parameters $\theta^{Q'}$ and $\theta^{\mu'}$.

In Algorithm~\ref{alg:DDPG}, the expectations in \eqref{eq:ddpg_actor_grad} and \eqref{eq:ddpg_critic_loss} are implemented by minibatch estimates over transitions sampled from the replay memory.

A critical component of DDPG is the use of target networks, which are time-delayed copies of the original actor and critic networks. These target networks are updated through soft updates to ensure stable learning:
\begin{align}
    \theta^{Q'} &\leftarrow \tau \theta^Q + (1 - \tau) \theta^{Q'}, \label{eq:ddpg_soft_q} \\
    \theta^{\mu'} &\leftarrow \tau \theta^\mu + (1 - \tau) \theta^{\mu'}, \label{eq:ddpg_soft_mu}
\end{align}
where $\tau \ll 1$ is a small constant that controls the update rate. This soft update mechanism helps to stabilise training by preventing the target values from changing too rapidly.

For exploration in continuous action spaces, DDPG uses an exploration policy that adds temporally correlated noise to the actor output:
\begin{equation}
    a_t = \mu(s_t|\theta^\mu) + \mathcal{N}_t,
\end{equation}
where $\mathcal{N}_t$ follows an Ornstein--Uhlenbeck process with reversion rate $\nu_{\mathrm{OU}}$ and noise scale $\sigma_{\mathrm{OU}}$, which generates temporally correlated exploration that is well-suited for physical control problems with inertia.

\begin{figure}[tb]
\centering
\includegraphics[width=0.85\linewidth]{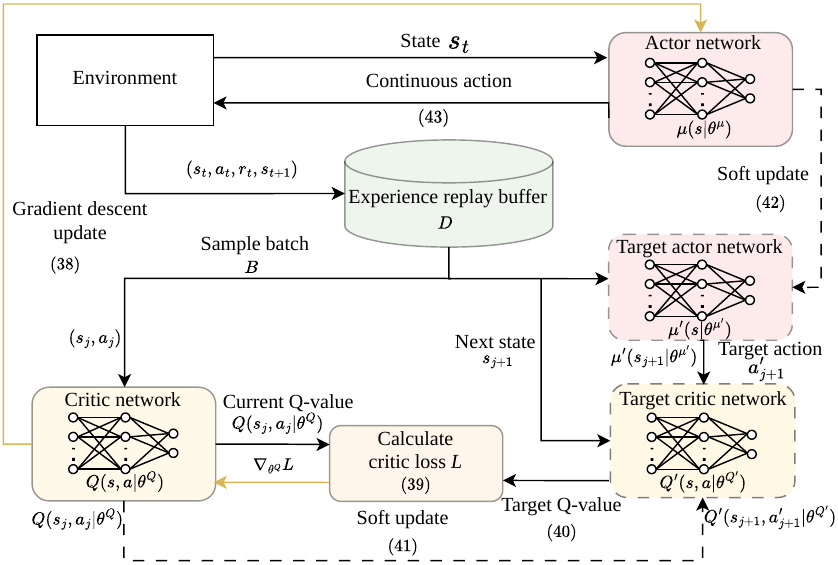}
\caption{Architecture of the DDPG algorithm.}
\label{fig:DDPG}
\end{figure}

The DDPG algorithm used in this study integrates the above techniques to provide stable and efficient learning for continuous energy management problems. The training procedure, which combines experience replay, actor-critic updates, and target networks, is summarised in Algorithm~\ref{alg:DDPG}, and its schematic is shown in Fig.~\ref{fig:DDPG}. In Algorithm~\ref{alg:DDPG}, $N_{\mathrm{ep}}$ denotes the total number of training episodes, $T_{\mathrm{total}}$ the number of time steps per episode, $N_{\mathrm{buf}}$ the replay-buffer capacity, and $B$ the minibatch size. The deployment procedure for evaluating the learned policy is presented in Algorithm~\ref{alg:ddpg_energy_management}.

\begin{algorithm}[tb]
\caption{DDPG training}
\label{alg:DDPG}
\begin{algorithmic}[1]
\STATE \textbf{Initialise:}
\STATE \quad Critic network $Q(s,a|\theta^Q)$ and actor network $\mu(s|\theta^\mu)$ with random weights
\STATE \quad Target networks: $\theta^{Q'} \leftarrow \theta^Q$,\; $\theta^{\mu'} \leftarrow \theta^\mu$
\STATE \quad Replay buffer $\mathcal{D}$ with capacity $N_{\mathrm{buf}}$; initialise OU noise process $\mathcal{N}$
\FOR{episode $= 1$ to $N_{\mathrm{ep}}$}
\STATE \quad Observe initial state $s_1$; reset noise $\mathcal{N}$
\FOR{$t = 1$ to $T_{\mathrm{total}}$}
\STATE \quad Select action: $a_t = \mu(s_t|\theta^\mu) + \mathcal{N}_t$
\STATE \quad Execute $a_t$; observe reward $r_t$, next state $s_{t+1}$, done flag $d_t$
\STATE \quad Store $(s_t, a_t, r_t, s_{t+1}, d_t)$ in $\mathcal{D}$
\IF{$|\mathcal{D}| \geq B$}
\STATE \quad Sample minibatch $\{(s_i, a_i, r_i, s_{i+1}, d_i)\}_{i=1}^{B}$ from $\mathcal{D}$
\STATE \quad Compute $y_i$ for each transition via \eqref{eq:ddpg_target}
\STATE \quad Update critic by minimising \eqref{eq:ddpg_critic_loss}
\STATE \quad Update actor via policy gradient \eqref{eq:ddpg_actor_grad}
\STATE \quad Soft-update target networks via \eqref{eq:ddpg_soft_q}--\eqref{eq:ddpg_soft_mu}
\ENDIF
\IF{$d_t = 1$}
\STATE \quad Terminate episode
\ELSE
\STATE \quad $s_t \leftarrow s_{t+1}$
\ENDIF
\ENDFOR
\ENDFOR
\end{algorithmic}
\end{algorithm}

\begin{algorithm}[tb]
\caption{DDPG-based energy management deployment}
\label{alg:ddpg_energy_management}
\begin{algorithmic}[1]
\REQUIRE Initial system state $s_1$; evaluation horizon $T_{\mathrm{total}}$; state normalisation function $\phi(\cdot)$
\ENSURE Continuous control action $a_t$ for each time slot
\STATE Load pretrained actor network weights $\theta^\mu$ from Algorithm~\ref{alg:DDPG}
\FOR{$t = 1$ to $T_{\mathrm{total}}$}
\STATE Normalise state: $\tilde{s}_t \gets \phi(s_t)$
\STATE Select action: $a_t \gets \mu(\tilde{s}_t\,|\,\theta^\mu)$
\STATE Execute $a_t$ in the industrial park environment
\STATE Observe next state $s_{t+1}$
\STATE $s_t \leftarrow s_{t+1}$
\ENDFOR
\end{algorithmic}
\end{algorithm}

\section{Results}\label{sec:Results}

This section evaluates the proposed DDPG-based energy management framework through simulations grounded in operational data from a real industrial park. We first describe the simulation setup and the baseline methods used for comparison. Subsequently, the training convergence, thermal comfort performance, scheduling behaviour, and overall cost reduction are analysed in detail to demonstrate the effectiveness of the proposed approach.

\subsection{Simulation Setup}

The load and PV data used in this study were collected from an EMS-equipped industrial park building in Fujian Province, China. The building primarily manufactures solar cables, with a daily peak production load of approximately 266.25~kWh. Data were recorded at hourly intervals over the period of 2024--2025. Time-of-use electricity prices are published in advance by the local power authority, and outdoor temperature profiles are obtained from the regional meteorological service. Because DR events are most frequent during the summer cooling season, September data are presented here for performance evaluation.

The DDPG agent is trained on the September 2024 dataset and subsequently evaluated on the September 2025 dataset to assess generalisation to unseen conditions. All system parameters are listed in Table~\ref{tab:parameters_system}, and the DRL hyper-parameters are given in Table~\ref{tab:parameters_DDPG}.

\begin{table}[tb]
\centering
\caption{System component parameters}\label{tab:parameters_system}
\begin{tabular}{lcc}
\toprule
Component & Parameter & Value \\
\midrule
\multirow{5}{*}{HVAC}
& $T_{\min}/T_{\max}$ & 20/24$^\circ$C \\
& $\eta^{\text{HVAC}}$ & 3.2 \\
& $G_{\text{b}}$ & 18 kW/$^\circ$C \\
& $P_{\max}^{\text{HVAC}}$ & 50 kW \\
& $\varepsilon$ & 0.85 \\
\addlinespace
\multirow{3}{*}{ESS \& EV}
& $\text{SoC}_{\min}/\text{SoC}_{\max}$ & 0.2/1.0 \\
& $P_{\text{rated}}$ & 100 kW \\
& $\eta_{\text{c}}/\eta_{\text{d}}$ & 0.95 \\
\addlinespace
\multirow{2}{*}{ESS}
& $E^{\text{ESS}}$ & 400 kWh \\
& Initial $\text{SoC}^{\text{ESS}}$ & 0.5 \\
\addlinespace
\multirow{3}{*}{EV}
& $\text{SoC}_{\text{lim}}$ & 0.6 \\
& $E^{\text{EV}}$ & 400 kWh \\
& Initial $\text{SoC}^{\text{EV}}$ & 0.35 \\
\addlinespace
\multirow{2}{*}{DR Policy}
& $L_{\text{dr}}$ & 100 kWh \\
& $p_{\text{dr}}$ & Dynamic \\
\addlinespace
\multirow{2}{*}{Carbon Emission}
& $\omega$ & 0.28088 kgCO$_2$/kWh \\
& $\rho_{\text{c}}$ & 6\% \\
\bottomrule
\end{tabular}
\end{table}

\begin{table}[tb]
\centering
\caption{Summary of DRL training settings for DDPG}\label{tab:parameters_DDPG}
\begin{tabular}{lc}
\toprule
Parameter & Value \\
\midrule
Actor and critic network layers & 2 \\
Actor and critic hidden nodes per layer & 256 \\
Actor hidden and output activation & ReLU, tanh \\
Critic hidden and output activation & ReLU, linear \\
\addlinespace
Actor learning rate & $1\times10^{-4}$ \\
Critic learning rate & $1\times10^{-3}$ \\
Optimiser & Adam \\
Batch size $B$ & 64 \\
Replay buffer size $N_{\mathrm{buf}}$ & $10^6$ \\
Discount factor $\gamma$ & 0.99 \\
Target update rate $\tau$ & 0.001 \\
Update frequency & Every step \\
\addlinespace
OU parameters $\nu_{\mathrm{OU}}, \sigma_{\mathrm{OU}}$ & 0.15, 0.2 \\
Training episodes $N_{\mathrm{ep}}$ & 7{,}500 \\
\bottomrule
\end{tabular}
\end{table}

\subsection{Baseline Methods}

Three rule-based strategies are introduced as baselines. All three employ the same HVAC control logic: the HVAC system cools when the indoor temperature exceeds the upper comfort bound and heats when it falls below the lower comfort bound.

\subsubsection{Baseline~1: Rule-based DR}
This strategy exploits the day-ahead DR dispatching signal. The ESS is fully discharged during the DR event window and charged to full capacity during off-peak hours. This approach captures DR revenue but does not optimise for time-of-use price arbitrage.

\subsubsection{Baseline~2: Time-of-use arbitrage}
This strategy charges the ESS during low-price periods and discharges it during high-price periods, following a conventional arbitrage logic. It does not explicitly participate in DR, however, if the high-price discharge windows coincide with DR event periods, incidental DR revenue may be obtained.

\subsubsection{Baseline~3: No storage dispatch}
This strategy neither participates in DR nor performs any active battery dispatch. It records the total electricity cost incurred by the industrial park under passive operation and serves as a reference for quantifying the value of active energy management.

\subsection{Performance Evaluation}

The evaluation is conducted on September 2025 data. For clarity, detailed scheduling profiles are presented for a representative 48-hour window, while cost comparisons are reported over the full evaluation month.

\subsubsection{Training Convergence}
\begin{figure}[tb]
\centering
\includegraphics[width=0.8\linewidth]{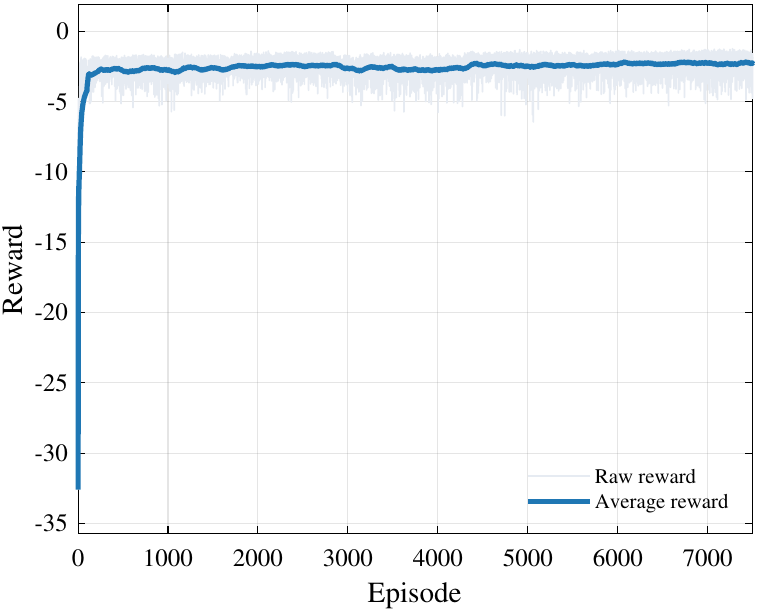}
\caption{Episode reward during the DDPG training process.}
\label{fig:rewards_DDPG}
\end{figure}

Fig.~\ref{fig:rewards_DDPG} illustrates the episode reward trajectory over 7{,}500 training episodes. The reward increases steadily during the early episodes and stabilises after approximately 5\,000 episodes, indicating that the agent has converged to a near-optimal policy. The final converged reward is approximately $-2$, where the negative sign reflects the net cost nature of the objective.

\begin{figure}[tb]
    \centering
    \subfloat[]{
        \includegraphics[width=0.9\linewidth]{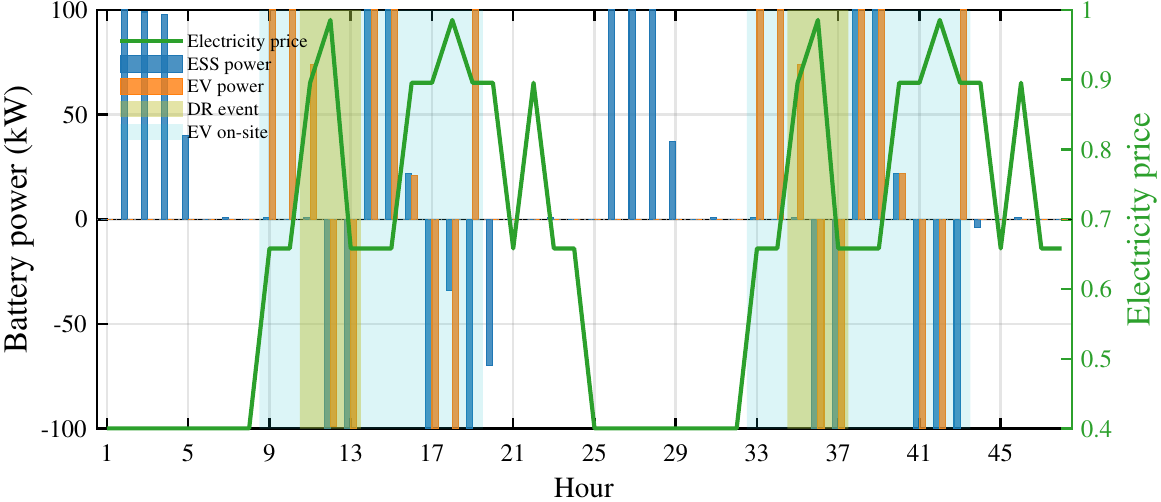}
        \label{fig:state_sub_a}
    }
    \hfill
    \subfloat[]{
        \includegraphics[width=0.9\linewidth]{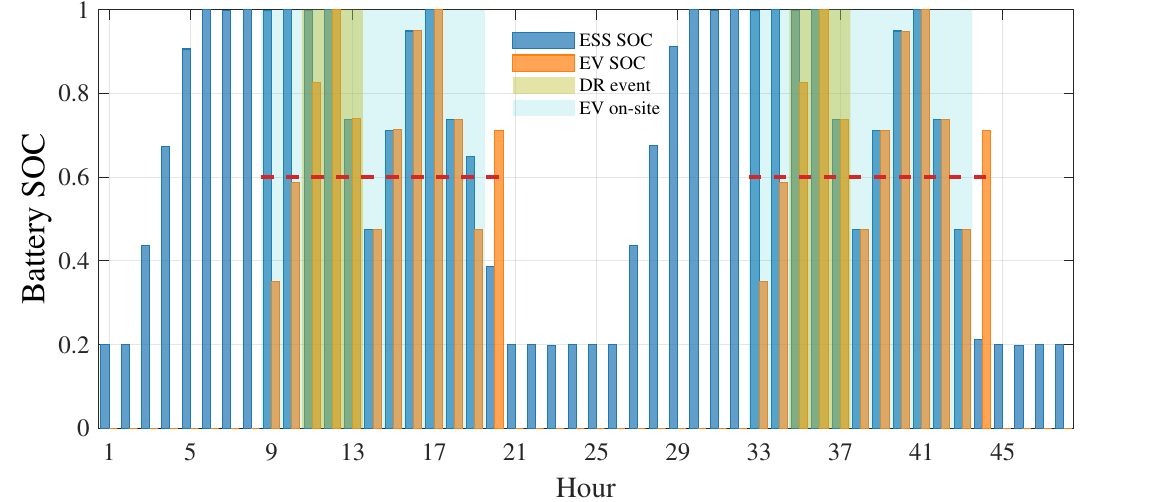}
        \label{fig:state_sub_b}
    }
    \hfill
    \subfloat[]{
        \includegraphics[width=0.9\linewidth]{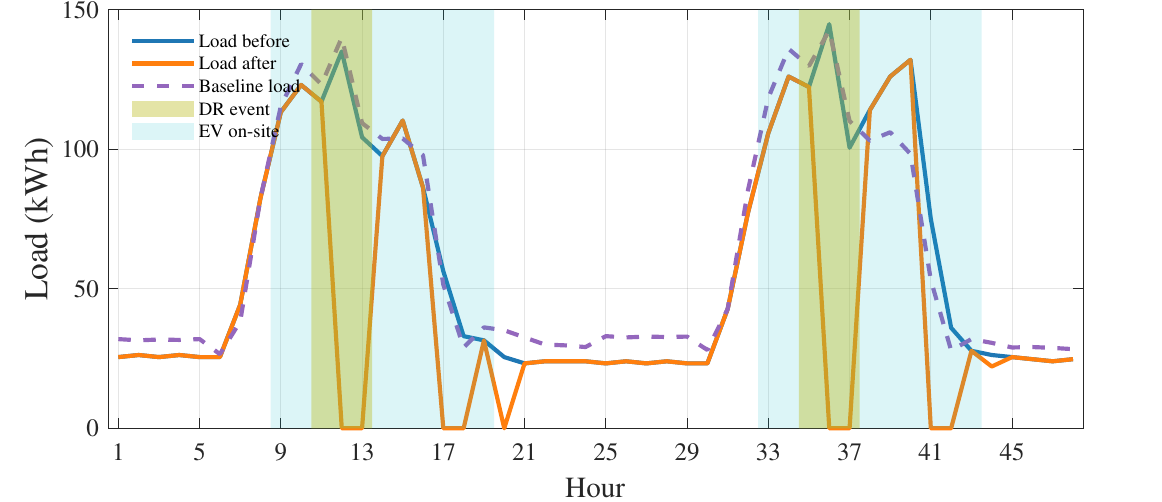}
        \label{fig:state_sub_c}
    }
    \caption{Detailed scheduling results: (a) charging and discharging power of the ESS and EV, where positive values denote charging and negative values denote discharging; (b) state of charge trajectories of both batteries; (c) production load before and after DR adjustment.}
    \label{fig:state}
\end{figure}

\subsubsection{Scheduling Analysis}

After training, the learned actor network is deployed with the September 2025 electricity prices, outdoor temperatures, and building loads. Fig.~\ref{fig:state} presents the detailed scheduling results for a representative 48-hour window from 9--10 September 2025. The green line indicates the time-of-use electricity price, the shaded green bands mark the DR event windows, and the shaded blue bands indicate the EV on-site periods.

\subsubsection{Thermal Comfort Performance}

\begin{figure}[tb]
\centering
\includegraphics[width=0.85\linewidth]{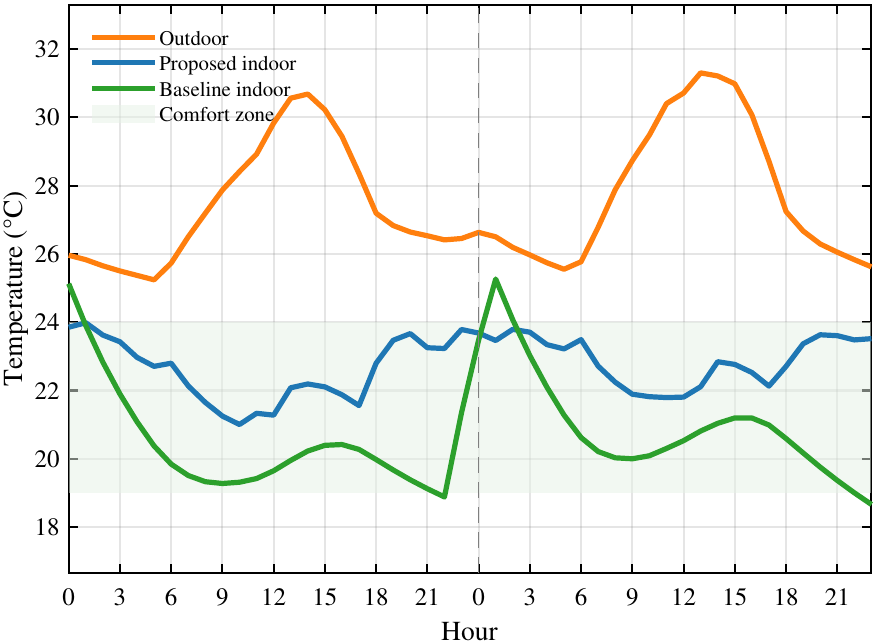}
\caption{Indoor and outdoor temperature profiles under the proposed DDPG-based control over a 48-hour period.}
\label{fig:temp_control}
\end{figure}

Maintaining the indoor temperature within the prescribed comfort zone is a key objective of the proposed framework. Fig.~\ref{fig:temp_control} depicts the outdoor and indoor temperature profiles under the proposed DDPG control over the 48-hour evaluation window. Despite considerable fluctuations in the outdoor temperature, the learned policy regulates the HVAC power to keep the indoor temperature within the comfort bounds of $[T_{\min},\, T_{\max}]$ throughout the period. This result confirms that the temperature deviation penalty incorporated in the reward function effectively guides the agent to satisfy the thermal comfort requirement.

Several observations can be drawn from the results. First, the proposed method learns to charge the batteries during low-price periods and discharge them during high-price periods, thereby performing effective energy arbitrage. Second, during DR event windows, both the ESS and EV are discharged to reduce the apparent grid consumption of the park. As shown in Fig.~\ref{fig:state}(c), the adjusted production load remains below the baseline load throughout the DR periods, fulfilling the response obligation and generating DR revenue. Third, the EV battery is managed to reach the departure SoC threshold before the vehicle leaves the site, satisfying the mobility constraint. These results demonstrate that the DDPG agent has learned a coordinated multi-objective scheduling strategy that balances DR participation, price arbitrage, and operational constraints.

\subsubsection{Cost Comparison}

\begin{figure}[tb]
\centering
\includegraphics[width=0.9\linewidth]{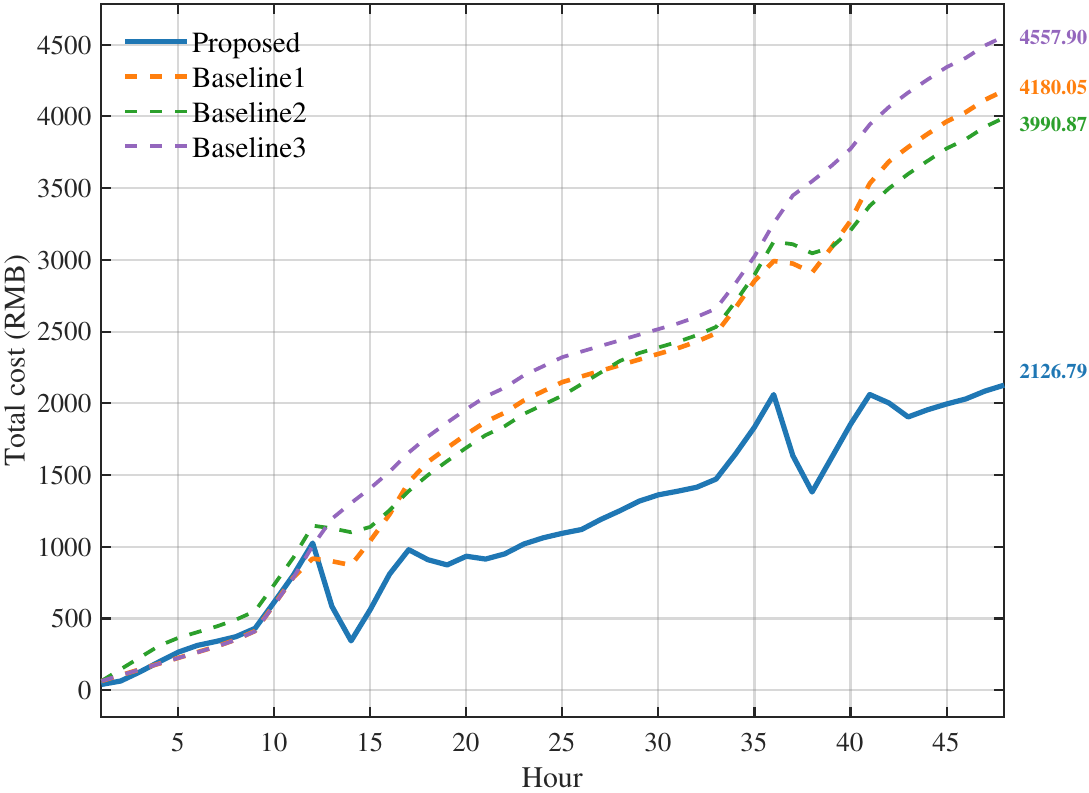}
\caption{Total operating costs over the evaluation window.}
\label{fig:profit}
\end{figure}

Fig.~\ref{fig:profit} and Table~\ref{tab:month_cost_compare} summarise the total operating costs of the proposed method and the three baseline strategies over the evaluation month. The proposed DDPG-based method achieves a total cost of 34{,}624.44~RMB, representing savings of 44.58\% compared with Baseline~1, 40.68\% compared with Baseline~2, and 58.89\% compared with Baseline~3.

\begin{table}[tb]
\renewcommand{\arraystretch}{1.3}
\caption{Total operating costs over the evaluation month}
\label{tab:month_cost_compare}
\centering
\begin{tabular}{l c c}
\hline
\multirow{2}{*}{Method} & \multicolumn{2}{c}{Total cost} \\
\cline{2-3}
 & Value (RMB) & Saving \\
\hline
Proposed &  34{,}624.44 & -- \\
Baseline~1 & 50{,}059.46 & 44.58\% \\
Baseline~2 & 48{,}708.06 & 40.68\% \\
Baseline~3 & 55{,}013.60 & 58.89\% \\
\hline
\end{tabular}
\end{table}

Evaluating the baseline models provides a clearer understanding of the cost dynamics. Baseline~2, which performs time-of-use arbitrage, achieves a lower cost than Baseline~1, which follows a rigid DR-focused discharge schedule without optimising for price differences. This indicates that conducting price arbitrage while doing DR response contributes more to cost reduction than DR revenue alone under the evaluated conditions. Note that Baseline~2 may also capture incidental DR revenue when its high-price discharge windows overlap with DR event periods. Baseline~2 outperforms Baseline~3 by approximately 12.9\%, confirming the value of active battery dispatch even without explicit DR participation. The proposed method surpasses all baselines because it jointly optimises DR participation, energy arbitrage, thermal comfort, and battery ageing through a unified learned policy, capturing coupled dynamics that rule-based strategies cannot exploit.

\section{Conclusion}\label{sec:Conclusion}

This paper presented a DDPG-based energy management framework for industrial park buildings equipped with PV systems, heterogeneous batteries, and diversified loads under a practical DR policy. The problem is formulated as an MDP that jointly optimises DR revenue, grid interaction cost, carbon emissions, battery ageing, thermal comfort, and EV departure SoC requirements. A dynamic energy distribution ratio captures the distinct characteristics of office and production zones, and dispatch-level ageing models are incorporated for both LFP-based ESS and NMC-based EV batteries. Simulation results on operational data from Fujian Province, China show that the proposed method maintains indoor thermal comfort and satisfies EV mobility needs whilst achieving cost reductions of 44.58\%, 40.68\%, and 58.89\% compared with a rule-based DR strategy, a time-of-use arbitrage approach, and a passive baseline, respectively.

\bibliographystyle{bib/IEEEtran}
\bibliography{bib/DR_ref}

\end{document}